\documentclass[pra,twocolumn,showpacs]{revtex4}
\usepackage{graphicx,bbm,amssymb,amsmath}
\begin{document}
\title{State reconstruction by on/off measurements}
\author{Alessia Allevi}
\affiliation{CNISM UdR Milano, I-20133 Milano, Italy}
\author{Alessandra Andreoni}
\affiliation{Dipartimento di Fisica e Matematica, Universit\`a
dell'Insubria, I-22100, Como, Italy}
\affiliation{CNISM UdR Como, I-22100 Como, Italy}
\author{Maria Bondani}\email{maria.bondani@uninsubria.it}
\affiliation{ULTRAS CNR-INFM, I-22100 Como, Italy}
\author{Giorgio Brida}
\affiliation{Istituto Nazionale di Ricerca Metrologica, I-10135, Torino, Italy}
\author{Marco Genovese}\email{m.genovese@inrim.it}
\affiliation{Istituto Nazionale di Ricerca Metrologica, I-10135, Torino, Italy}
\author{Marco Gramegna}
\affiliation{Istituto Nazionale di Ricerca Metrologica, I-10135, Torino, Italy}
\author{Stefano Olivares}
\affiliation{CNISM UdR Milano, I-20133 Milano, Italy}
\affiliation{Dipartimento di Fisica, Universit\`a degli Studi di Milano, I-20133 Milano, Italy}
\author{Matteo G. A. Paris}\email{matteo.paris@fisica.unimi.it}
\affiliation{Dipartimento di Fisica, Universit\`a degli Studi di Milano, I-20133 Milano, Italy}
\affiliation{CNISM UdR Milano, I-20133 Milano, Italy}
\affiliation{ISI Foundation, I-10133 Torino, Italy}
\author{Paolo Traina}
\affiliation{Istituto Nazionale di Ricerca Metrologica, I-10135, Torino, Italy}
\author{Guido Zambra}
\affiliation{Dipartimento di Elettronica, Politecnico di Milano, I-20133 Milano, Italy}
\begin{abstract}
We demonstrate a state reconstruction technique which provides either the
Wigner function or the density matrix of a field mode and requires only
avalanche photodetectors, without any phase or amplitude discrimination
power. It represents an alternative to
quantum homodyne tomography of simpler implementation.
\end{abstract}
\maketitle
\section{Introduction}
The characterization of states and operations at the quantum level plays
a leading role in the development of quantum technology.
A state reconstruction technique is a method that provides the
complete description of a physical system upon the measurements of 
an observable or a set of observables \cite{gentomo}. An effective 
reconstruction technique gives the maximum possible knowledge of the 
state, thus allowing one to make the best, at least the best probabilistic, 
predictions on the results of any measurement that may be performed on 
the system.  
At a first sight,
there is an unavoidable tradeoff between the complexity of the detection
scheme and the amount of extractable information, which can be used
to reconstruct the quantum state \cite{mg}.  Currently, the most
effective quantum state reconstruction technique for the radiation field
is quantum homodyne tomography \cite{wv99,sp}, which requires the
measurement of a continuous set of field quadrature and allows for the
reliable reconstruction of any quantity expressible in terms of an
expectation value \cite{rec1,rec2,rec3,rec4,rec5,rec6}.  
A question arises on whether the
tradeoff may be overcome by a suitable experimental configuration or it
corresponds to some fundamental limitations.  Here we demonstrate that
no specific discrimination power is required to the detector in either
amplitude or phase, and that full state reconstruction is possible by a
suitable processing of the data  obtained with detectors revealing light
in the simplest way, i.e. on/off detectors, such as single-photon
avalanche photodiodes.  Of course, some form of phase and/or amplitude
modulation is necessary, which, in our scheme, is imposed to the field
before the detection stage.  In fact, our technique is built on the
completeness of any set of displaced number states
\cite{WB96,WV96,L96,TO971,TO972,KB99} and the reliable maximum likelihood
reconstruction of arbitrary photon-number distributions \cite{CVP} from
on/off data. 
\par
The paper is structured as follows. In Section \ref{s:rec} we describe
our reconstruction method, whereas in Section \ref{s:exp} the
experimental setup used in the reconstruction is described in some
details. Results are illustrated in Section \ref{s:res} and the error
analysis is reported in Section \ref{s:err}. In Section \ref{s:dis} we 
discuss few additional topics while Section \ref{s:out} closes
the paper with some concluding remarks.
\section{State reconstruction by on/off measurements} \label{s:rec}
We start to describe our reconstruction technique by observing that the
modulation of a given signal, described by the density matrix $\varrho$,
corresponds to the application of a coherent displacement (probe)
$\varrho_\alpha = D(\alpha) \varrho D^\dag (\alpha)$, $\alpha\in
{\mathbb C}$. In practice, it can be easily obtained by mixing the state
under investigation with a known coherent reference in a beam-splitter
or a Mach-Zehnder interferometer \cite{bs96}.
Upon varying
amplitude and phase of the coherent reference and/or the overall transmissivity of
the interferometer, the modulation may be tuned in a relatively broad range
of values.  The main idea behind our method is simple: the photon
distributions of coherently modulated signals, i.e. the diagonal
elements $p_n(\alpha)=\langle n |\varrho_\alpha |n\rangle$ of the
density matrix $\varrho_\alpha$, contain relevant information about the
complete density matrix of the original signal $\varrho$.  Upon
measuring or reconstructing the photon distribution $p_n(\alpha)$ for
different values of the modulation one has enough information for full
state reconstruction. By re-writing the above relation as
$p_n(\alpha) = \sum_{km} D_{nk}(\alpha) \varrho_{km} D_{mn} (\alpha)$,
the off diagonal matrix elements may be recovered upon inversion by
least square method, i.e. \cite{TO971}
$$\langle m+s | \varrho |m\rangle = N_\phi^{-1} \sum_{l=1}^{N_\phi}
\sum_{n=0}^{\bar n} F^{(s)}_{nm} (|\alpha|) p_n (|\alpha|
e^{i \phi_l}) e^{i s \phi_l}$$
where $N_\phi$ is the number of modulating phases, $\bar n$ the
truncation dimension of the Fock space, and $F$ depends only on
$|\alpha|$ \cite{TO971}. State reconstruction by the above formula
requires, in principle, only phase modulation of the signal under
investigation.  Maximum likelihood methods and iterative procedures may
be also used \cite{ZJ06}.  On the other hand, the Wigner function may be
reconstructed using its very definition in terms of displacement 
\cite{Cahill}
$$W(\alpha) = \hbox{Tr}[D(\alpha)\varrho D^\dag (\alpha)\, (-)^{a^\dag
a}] = \sum_n (-)^n\, p_n (\alpha)\:.$$ 
As a matter of fact, the measurement of the photon distribution is
challenging as photo-detectors that can operate as photon counters are
rather rare and affected either by a low quantum efficiency \cite{burle}
or require cryogenic conditions, thus impairing common use \cite{xxx,
serg}. Therefore, a method with displacement
but without photo-counting has been used so far only for states in 
the 0-1 subspace of the Fock space \cite{KB99}.
On the other hand, the experimental reconstructions of
photon-number distributions for both continuous-wave and pulsed light
beams is possible using simple on/off single-photon avalanche
photodetectors.
This requires the collection of the frequencies of the {\em off} events,
$P_{0,k}=\sum_{n = 0}^\infty{( {1 - \eta_k})^n p_n}$ at different
quantum efficiencies of the detector, $\eta_k$. The data are then used
in a recursive maximum likelihood reconstruction algorithm that yields
the photon-number distributions as $$p_n^{i + 1}  = p_n^i \sum_{k = 1}^K
(A_{kn}/\sum_j {A_{kj}})(P_{0,k}/p_{0,k}[\{p_n^i\}]),$$ where
$A_{kn}  = \left({1 - \eta _k } \right)^n$ and $p_{0,k}$ is the
probability of {\em off} events calculated from the reconstructed
distribution at the $i$th iteration \cite{pcount}. The effectiveness of
the method has been demonstrated for single-mode \cite{CVP} and
multimode fields \cite{bip}, and also applied to improve quantum key
distribution \cite{TM08}.
\par
Since the implementation of the modulation is relatively easy,
we have thus a reconstruction technique which
provides the quantum state of radiation modes and requires only
avalanche detectors, without any phase or amplitude discrimination
power. Here, we develop the idea into a proper
reconstruction technique and demonstrate the reconstruction of the
Wigner function \cite{silb} and the density matrix for
different states of the optical field.
\section{Experimental setups}\label{s:exp}
\begin{figure}[h!]
\includegraphics[width=0.8\columnwidth]{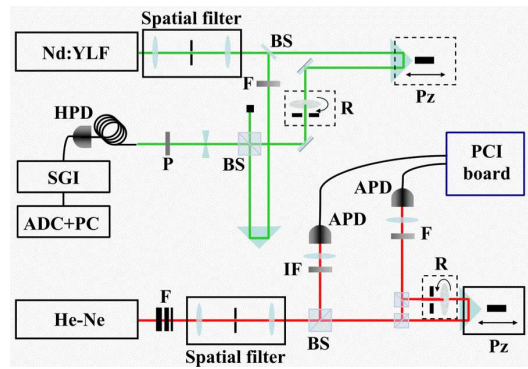}
\caption{(Color online) Schematic diagram of the two experimental
setups. Upper panel: Nd:YLF, pulsed laser source; P, polarizer; HPD,
hybrid photodetector; SGI, synchronous gated integrator; ADC,
analog-to-digital converter. Lower panel: He-Ne, cw laser source; IF,
interference filter; APD, single-photon avalanche photodiode. F,
neutral-density filter; BS, beam splitter; Pz, piezoelectric movement;
R, rotating ground glass plate.  Components in dotted boxes are
inserted/activated when necessary.} \label{f:setup}
\end{figure}

We have performed two experiments for the reconstruction of the Wigner
function and the density matrix respectively.  In Fig.~\ref{f:setup} we
sketch the corresponding experimental setups: the upper panel for the
measurement of the Wigner function and the lower panel for the density
matrix (lower panel).  The first set of measurements was performed on
ps-pulsed light fields at 523 nm wavelength. The light source was the
second-harmonic output of a Nd:YLF mode-locked laser amplified at 5 kHz
(High Q Laser Production) delivering pulses of $\sim 5.4$~ps duration.
\par
The field, spatially filtered by means of a confocal telescope, was
split into two parts to give both signal and probe.  The photon-number
distribution of the probe was kept Poissonian, while the coherent
photon-number distribution of the signal field was modified in order to
get suitable states of light. In particular, we have generated two
phase-insensitive classical states, namely a phase-averaged coherent
state and a single-mode thermal state. The first one was obtained by
changing the relative phase between signal and probe fields at a
frequency of $\sim300$ Hz by means of a piezoelectric movement (Pz in
the upper panel of Fig.~\ref{f:setup}), covering 1.28~$\mu$m span. On
the other hand, to get the single-mode thermal state we inserted an
Arecchi's rotating ground glass disk on the pathway of the signal field
followed by a pin-hole that selected a single coherence area.  
\par
Signal and probe fields were mixed in an unpolarizing cube beam-splitter
(BS) and a portion of the exiting field was sent, through a multimode
optical fiber (600 $\mu$m core-diameter), to a hybrid photodetector
module (HPD, mod.  H8236-40, Hamamatsu, maximum quantum efficiency
$\eta_\mathrm{HPD}$=0.4 at 550 nm). Although the detector is endowed
with partial photon resolving capability, we used it as a simple on/off
detector by setting a threshold at the value corresponding to zero
detected photons. Its output current pulses were suitably
gate-integrated by a SR250 module (Stanford Research Systems, CA) and
sampled to produce a voltage, which was digitized and recorded at each
shot.  In order to modulate the amplitude of the probe field, a variable
neutral density filter was placed on its pathway. The maximum overall
detection efficiency, calculated by including the losses of the
collection optics, was $\eta_{max}=0.29$.  We used a polarizer put in
front of the fiber to vary the quantum efficiency of the detection chain
from $\eta_{max}$ to 0.
\section{Results}\label{s:res}
Here we illustrate the reconstruction obtained for the Wigner function
and the density matrix of phase-averaged coherent states and thermal
states, as obtained by our method after recording the on/off statistics
of amplitude- and/or phase-modulated signals.
\begin{figure}[h!]
\includegraphics[width=0.88\columnwidth]{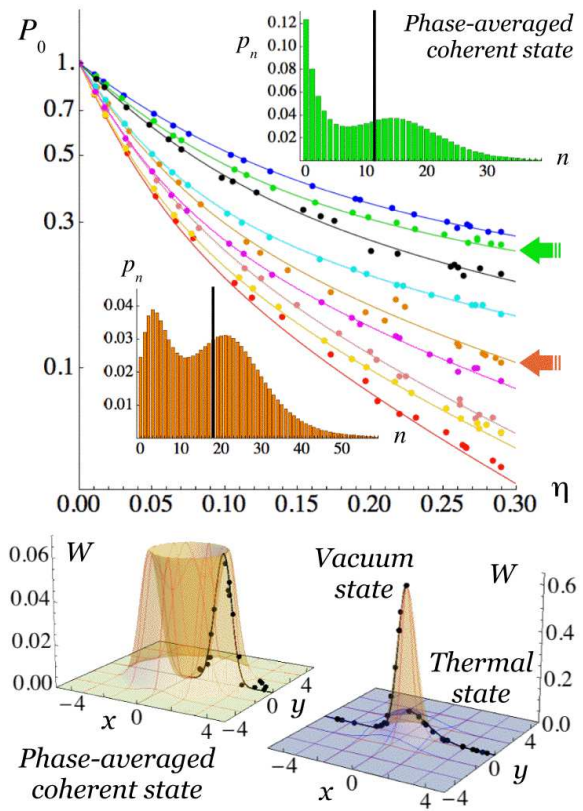}
\caption{(Color online) State reconstruction by amplitude-modulation and
on/off measurements. In the main plot we report the {\em off} frequencies
as a function of the quantum efficiency as obtained when the signal under
investigation is a phase-averaged coherent state and for different
values of the probe intensity $|\alpha|^2$. The two insets show the
reconstructed photon distributions for $|\alpha|^2=5.02$ and
$|\alpha|^2=10.69$, corresponding to the off distributions indicated by
the arrows.  The vertical black bars denote the mean value of the photon
number for the two distributions ($\langle a^\dag a\rangle = 11.3$,
upper,  and $\langle a^\dag a\rangle = 18.0$, lower).  In the lower
left plot we report the corresponding reconstructed Wigner function.
In the lower right plot we report the Wigner functions for
signals in (blue) thermal state and (yellow, with sharper peak) vacuum .
\label{f:wsamm}}
\end{figure}
\par
In Fig.~\ref{f:wsamm} we report the reconstructed Wigner functions for a
phase-averaged coherent state with real amplitude $z \simeq2.1$ and a
thermal state with average number of photons $n_{th}\simeq2.4$.
The Wigner function of the vacuum state
is also reported for comparison. As it is apparent from the plots all
the relevant features of the Wigner functions are well recovered,
including oscillations and the broadening due to thermal noise.  In this
case raw data are frequencies of the {\em off} event (collected over 30000 laser shots)
as a function of
detector efficiency, taken at different amplitudes of the modulating
field, whereas the intermediate step corresponds to the reconstruction
of the $p_n(\alpha)$. The insets of Fig.~\ref{f:wsamm} display $p_n(\alpha)$
for the phase-averaged coherent state at
two values of the modulation.
\begin{figure}[h!]
\includegraphics[width=0.88\columnwidth]{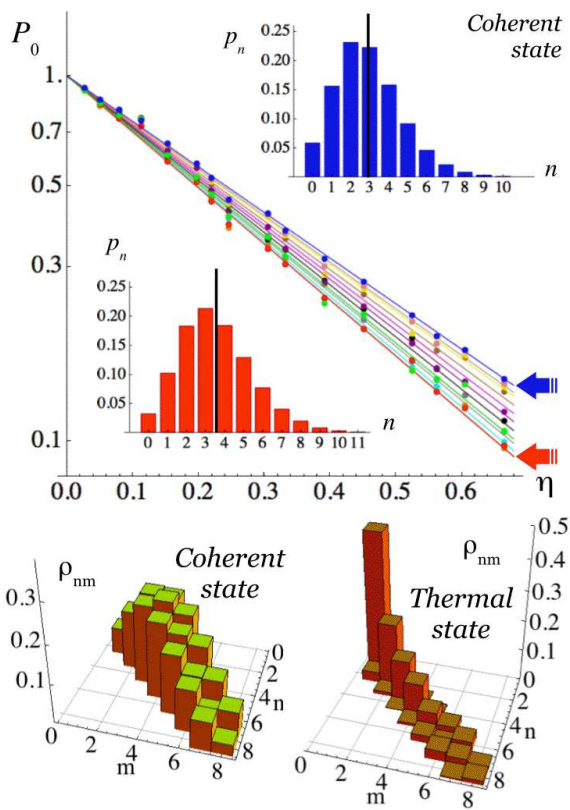}
\caption{(Color online) State reconstruction by phase-modulation and
on/off measurements. In the upper plot we report the {\em off}
frequencies as a function of the quantum efficiency as obtained when the
signal under investigation is a coherent state and for different
phase-shifts. The two insets show the reconstructed photon distributions
for the two phase-modulated versions of the signal corresponding to
maximum and minimum visibility at the output of the Mach-Zehnder
interferometer.  The vertical black bars denote the mean value of the
photon number for the two distributions, $\langle a^\dag a\rangle = 3.5$  and
$\langle a^\dag a\rangle = 2.9$.  In the lower left plot (left)
we report the corresponding reconstructed density matrix in the Fock
representation (diagonal and subdiagonal elements). In the lower right
plot we report the density matrix for the signal excited in a thermal
state.
\label{f:dmphm}}
\end{figure}
\par
The second set of measurements has been performed to achieve state
reconstruction with phase modulation. Here the light source was a He-Ne laser
beam attenuated to single-photon regime by neutral density filters. The spatial
profile of the beam was purified from non-Gaussian components by a
spatial filter. Beyond a beam-splitter, part of the beam was addressed
to a control detector in order to monitor the laser amplitude
fluctuations, while the remaining part was sent to a Mach-Zehnder
interferometer. A piezo-movement system allowed changing the phase
between the ``short'' and ``long'' paths by driving the position of the
reflecting prism with nanometric resolution and high stability. The beam
on the short arm represented the probe, while the beam on the long arm
was the state to be reconstructed. In the first acquisition the signal
was the coherent state itself while in the second acquisition it was a
pseudo-thermal state generated as described above. The detector, a Perkin-Elmer
Single Photon Avalanche Photodiode (SPCM-AQR) with quantum efficiency
$\eta_{max}=0.67$, was gated by a 20 ns time window at a
repetition rate of 200~kHz. To obtain a reasonable statistics,
a single run consisted of 5 repetitions of acquisitions lasting 4~s each.
In the bottom part of Fig.~\ref{f:dmphm} we report the reconstructed
density matrix in the
Fock representation (diagonal and subdiagonal) for a coherent state
with real amplitude $z\simeq1.8$ and a thermal state with
average number of photons equal to $n_{th}\simeq1.4$.
\par
As it is apparent from the plots the off-diagonal elements are correctly
reproduced in both cases despite the limited visibility.  Here the raw
data are frequencies of the {\em off} event as a function of the
detector efficiency, taken at different phase modulations, $\phi$,
whereas the intermediate step corresponds to the reconstruction of the
photon distribution for the phase-modulated signals. In our experiments
we used $N_\phi=12$ and $|\alpha|^2=0.01$ for the coherent state and
$|\alpha|^2=1.77$ for the thermal state.  The use of a larger $N_\phi$
would allow the reliable reconstruction of far off-diagonal elements,
which is not possible in the present configuration.  In the insets of
Fig.~\ref{f:dmphm} we report the reconstructed distributions at the
minimum and maximum of the interference fringes.
\section{Error analysis} \label{s:err}
\begin{figure}[h!]
\includegraphics[width=0.88\columnwidth]{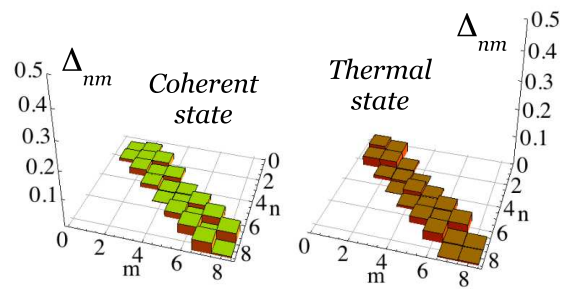}
\caption{(Color online) Absolute difference $\Delta_{nm}=|\varrho^{exp}_{nm}
-\varrho^{th}_{nm}|$ between reconstructed and theoretical values
of the density matrix elements for the coherent (left) and thermal
(right) states used in our experiments.}\label{f:err}
\end{figure}

The evaluation of uncertainties on the reconstructed states involves the
contributions of experimental fluctuations of on/off frequencies as well
as the statistical fluctuations connected with photon-number
reconstruction.  It has been argued \cite{KER97,KER99} that fluctuations
involved in the reconstruction of the photon distribution may generally
result in substantial limitations in the information obtainable on
the quantum state, e.g. in the case of multipeaked distributions \cite{zz}.
For our purposes this implies
that neither large displacement amplitudes may be employed, nor states
with large field and/or energy may be reliably reconstructed, although
the mean values of the fields measured here are definitely
non-negligible.  On the other hand, for the relevant regime of weak
field or low energy, observables characterizing the quantum state can be
safely evaluated from experimental data, including e.g. the parity
operator used to reconstruct the Wigner function in the phase-space. In
our experiments, the errors on the reconstructed Wigner function are of
the order of the size of the symbols in Fig. \ref{f:dmphm} whereas the
absolute errors $\Delta_{nm}=|\varrho^{exp}_{nm}-\varrho^{th}_{nm}|$ on
the reconstruction of the density matrix in the Fock basis are reported
in Fig. \ref{f:err}. 
\section{Discussions}\label{s:dis}
We have so far reconstructed the Wigner function and the density matrix
for coherent and thermal states. The extension to highly nonclassical
states does not require qualitative changes in the setups. The only
difference stays in the displacement, which should be obtained with high
transmissivity beam splitter in order to avoid mixing of the signal
\cite{bs96}.  \par 
As our method involves a beam splitter where the signal interferes with
a reference state in order to obtain the displacement, we have optimized
the effectiveness of mode matching and of the overall scheme by standard
visibility test.  We note that in this point our technique is similar to
quantum homodyne tomography (QHT), where the signal is amplified by the
mixing at a beam splitter with a strong local oscillator. The main
difference with standard QHT, however, is the spectral domain of the
measurement, which for QHT is confined to the sideband chosen for the
measurement, while it is not in our case.  The use of pulsed temporal
homodyning \cite{ht1,ht1a,ht2,ht3,ht4} would remove this limitation of
QHT. However, this technique is still challenging from the experimental
point of view and thus of limited use.  The effect of photodetectors
efficiency should be also taken into account. This  is a crucial issue
for QHT, which may even prevent effective reconstruction \cite{prec}. 
For the present on/off reconstrucion, it does not dramatically affect 
the accuracy \cite{pcount}.
Notice also that any uncertainty in the nominal
efficiency $\eta_{max}$ of the involved photodetectors does not
substantially affect the reconstruction \cite{pcount}.
\par
We stress that our method is especially suited for low excited states,
as it does not involve intense fields and delicate balancing to reveal 
the relevant quantum fluctuations. 
\section{Conclusions}\label{s:out}
In conclusion, we have demonstrated a state reconstruction technique
providing Wigner function and density matrix of a field mode starting
from on/off photodetection of amplitude- and/or phase-modulated versions
of the signal under investigation. Our scheme is little demanding as to
the detectors, with the amplitude and phase control required for full
state characterization transferred to the optical setup, and
appears to be reliable and simple especially for states with
small number of photons. 
We foresee a possible widespread use in emerging quantum technologies 
like quantum information, metrology and lithography.
\section*{Acknowledgments}
This work has been partially supported by the CNR-CNISM agreement,
EU project QuCandela, Compagnia di San Paolo, MIUR-PRIN-2007FYETBY
(CCQOTS), and Regione Piemonte (E14).


\begin{thebibliography}{99}
\bibitem{gentomo} G. M. D'Ariano, L. Maccone, and M. G. A. Paris,
J. Phys. A, {\bf 34}, 93 (2001).
\bibitem{mg} Y. Bogdanov et al., JETP Lett. {\bf 82}, 180 (2005);
C. Marquardt et al., Phys. Rev. Lett. {\bf 99}, 220401 (2007).
\bibitem{wv99} D. G. Welsch et al., Prog. Opt. {\bf 39}, 63 (1999).
\bibitem{sp} {\em Quantum State Estimation}, Lect. Not. Phys. {\bf 649},
M. G. A. Paris and  J. Rehacek Eds. (Springer, Berlin, 2004);
A. I. Lvovsky and M. G. Raymer, Rev. Mod. Phys. {\bf 81}, 299 (2009).
\bibitem{rec1}
D.T. Smithey et al., Phys. Rev. Lett. {\bf 70}, 1244 (1993);
\bibitem{rec2}
G. Breitenbach et al., Nature {\bf 387}, 471 (1997);
\bibitem{rec3}
M. Vasilyev et al., Phys. Rev. Lett. {\bf 84}, 2354 (2000);
\bibitem{rec4}
I. Lvovsky et al., Phys. Rev. Lett. {\bf 87}, 050402 (2001);
\bibitem{rec5}
V. Parigi et al., Science {\bf 317}, 1891 (2008);
\bibitem{rec6}
V. D'Auria et al., Phys. Rev. Lett. {\bf 102}, 020502 (2009).
\bibitem{WV96} S. Wallentowitz and W. Vogel, Phys. Rev. A {\bf 53}, 4528 (1996).
\bibitem{WB96} K. Banaszek and K. Wodkiewicz, Phys. Rev. Lett. {\bf 76}, 4344 (1996).
\bibitem{L96} D. Leibfried et al., Phys. Rev. Lett. {\bf 77}, 4281 (1996).
\bibitem{TO971} T. Opatrny and D. G. Welsch, Phys. Rev A {\bf 55}, 1462 (1997).
\bibitem{TO972} T. Opatrny, D. G. Welsch, and W. Vogel, Phys. Rev A {\bf
56}, 1788 (1997).
\bibitem{KB99} K. Banaszek et al.,
Phys. Rev. A {\bf 60}, 674 (1999); Act. Phys. Slov. {\bf 49}, 643 (1999).
\bibitem{CVP} G. Zambra et al., Phys. Rev. Lett. {\bf 95}, 063602 (2005);
G. Brida et al., Las. Phys. {\bf 16}, 385 (2006); G. Brida et al., Open
Syst. \& Inf. Dyn. {\bf 13}, 333 (2006); M. Bondani et al., 
Adv. Sci. Lett. (in press) ArXiv:0810.4026.
\bibitem{bs96} M. G. A. Paris, Phys. Lett. A {\bf 217}, 78 (1996).
\bibitem{ZJ06} Z.~Hradil, D.~Mogilevtsev, and J.~\v{R}eh\'{a}\v{c}ek,
Phys. Rev. Lett. {\bf 96}, 230401 (2006);
New J. Phys. {\bf 10}, 043022 (2008).
\bibitem{Cahill} K. E. Cahill and R. J. Glauber, Phys. Rev. \textbf{177}, 1882 (1969).
\bibitem{burle} G. Zambra, M. Bondani, A. S. Spinelli, and A. Andreoni,
Rev. Sci. Instrum. {\bf 75}, 2762 (2004).
\bibitem{xxx} J. Kim, S. Takeuchi, Y. Yamamoto, and H.H. Hogue, Appl.
Phys. Lett. {\bf 74}, 902 (1999); A. Peacock et al., Nature {\bf
381}, 135 (1996); A.E.Lita et al., Opt. Exp. 16 (2008) 3032.
\bibitem{serg} G. Di Giuseppe, A. V. Sergienko, B. E. A. Saleh, and
M. C. Teich in {\em Quantum Information and Computation}, E. Donkor,
A. R. Pirich, and H. E. Brandt Eds., Proceedings of the SPIE {\bf
5105}, 39 (2003).
\bibitem{pcount} A. R. Rossi, S. Olivares, and M. G. A. Paris,
Phys. Rev. A {\bf 70}, 055801 (2004).
\bibitem{bip} G. Brida et al., Opt. Lett. {\bf 31}, 3508 (2006).
\bibitem{TM08} T. Moroder et al., ArXiv:0811.0027v1
\bibitem{silb} K. Laiho, M. Avenhaus, K. N. Cassemiro, Ch. Silberhorn,
New J. Phys. {\bf 11},  043012 (2009).
\bibitem{KER97} K. Banaszek and K. Wodkiewicz, J. Mod. Opt. {\bf 44}, 2441
(1997).
\bibitem{KER99} K. Banaszek, J. Mod. Opt. {\bf 46}, 675 (1999).
\bibitem{zz} G. Zambra and M. G. A. Paris, Phys. Rev. A {\bf 74}, 063830 (2006).
\bibitem{ht1} H. Hansen, T. Aichele, C. Hettich, P. Lodahl, A. I.
Lvovsky, J. Mlynek, S. Schiller, Opt. Lett. {\bf 26}, 1714 (2001).
\bibitem{ht1a} A. Zavatta, M. Bellini,P. L.  Ramazza, F. Marin, F. T.
Arecchi, J. Opt. Soc. Am. B {\bf 19}, 1189 (2002).
\bibitem{ht2} T. Hirano, H. Yamanaka, M. Ashikaga, I. Konishi, R.
Namiki, Phys. Rev. A{68}, 042331 (2003).
\bibitem{ht3} J. Wenger, R. Tualle-Brouri, P. Grangier, Opt. Lett. {\bf
29} 1267 (2004).
\bibitem{ht4} Y. Eto, T. Tajima, Y. Zhang Y, T. Hirano, Opt. Lett. {\bf
32}, 1698 (2007).
\bibitem{prec} G. M. D'Ariano, M. G. A Paris, and M. F. Sacchi, 
Adv. Im. Electr. Phys. {\bf 128}, 205 (2003). 
\end{thebibliography}
\end{document}